# Solution of the relativistic Dirac-Hulthén problem

A. D. Alhaidari

*Physics Department, King Fahd University of Petroleum & Minerals, Box 5047, Dhahran 31261, Saudi Arabia*
email: haidari@mailaps.org

The one-particle three-dimensional Dirac equation with spherical symmetry is solved for the Hulthén potential. The s-wave relativistic energy spectrum and two-component spinor wavefunctions are obtained analytically. Conforming to the standard feature of the relativistic problem, the solution space splits into two distinct subspaces depending on the sign of a fundamental parameter in the problem. Unique and interesting properties of the energy spectrum are pointed out and illustrated graphically for several values of the physical parameters. The square integrable two-component wavefunctions are written in terms of the Jacobi polynomials. The nonrelativistic limit reproduces the well-known nonrelativistic energy spectrum and results in Schrödinger equation with a "generalized" three-parameter Hulthén potential, which is the sum of the original Hulthén potential and its square.

## I. INTRODUCTION

The Hulthén potential [1] is of considerable significance to various applications in many areas of physics. This includes applications in nuclear and particle physics, atomic physics, condensed matter, and chemical physics. One may consult, for example, the literature cited in the papers of Ref. [2]. For spherically symmetric interaction, this potential has the general form $(1-e^{\omega r})^{-1}$, where $\omega$ is a real positive parameter. A well-known application of this potential in atomic and nuclear physics is the screened Coulomb potential. In this case it is written as $V(r) = -\frac{\omega Z e^{-\omega r}}{1-e^{-\omega r}}$, where $Z$ is the Coulomb charge and $\omega$ is the screening parameter. For small values of the screening parameter it is approximated as $V(r) \approx -\frac{Z}{r}e^{-\omega r}$, showing clearly the screening effect. The nonrelativistic Hulthén problem has a closed form analytic solution only for s-wave ($\ell = 0$) [3]. Several techniques were used to obtain approximate solutions in the case where the angular momentum is not zero [2,4,5]. Various methods were employed in obtaining the solution (energy spectrum and wave function) of the nonrelativistic Hulthén problem. Super-symmetric quantum mechanics, shape invariance, path integration, and dynamical group are four methods among many which were used in the search for exact solutions of the wave equation with the Hulthén potential.

The relativistic problem, on the other hand, did not receive adequate attention. Most of the limited work on this problem was in the context of the Klein-Gordon equation. The Klein-Gordon equation with the Hulthén potential was treated by Znojil [6]. The same problem but with both vector and scalar Hulthén-type potentials was later investigated more thoroughly by Domínguez-Adame [7]. The Green's function for the Klein-Gordon operator with these two potentials was obtained using path integral approach by Chetouani *et al.* [8]. The scattering state solutions of the s-wave Klein-Gordon equation with vector and scalar Hulthén potentials were obtained for regular and irregular boundary conditions by Talukdar *et al.* [9]. On the other hand, only very few articles were written on the Dirac equation with the Hulthén potential. Roy and Roychoudhury used the algebraic approach in tackling this problem [10]. They considered the Dirac equation with scalar and vector potentials with the space component of the later being zero. Both potentials are of the same Hulthén type but of different strengths. This made the potential in the Dirac equation looks like $\begin{pmatrix} V+S & 0 \\ 0 & V-S \end{pmatrix}$, where $V$ is the time component of the vector potential and $S$ is the scalar. Explicit results were obtained for $S = V$ making the potential in the Dirac equation of the form $\begin{pmatrix} 2V & 0 \\ 0 & 0 \end{pmatrix}$ and resulting in Schrödinger-like equation for the upper spinor component. However, it is believed that such a potential structure might result in a singular behavior of the solution. A remedy could be found in the introduction of an additional pseudo-scalar potential $W$ such that the overall potential structure in the Dirac equation becomes $\begin{pmatrix} V+S & W \\ W & V-S \end{pmatrix}$. Thus, taking $S = V$ results in the more regular potential form $\begin{pmatrix} 2V & W \\ W & 0 \end{pmatrix}$ while at the same time giving the sought after Schrödinger-like equation. Quite recently, Guo *et al.* presented a more appropriate treatment of the relativistic s-wave problem in the Dirac equation with non-minimal coupling to spherically symmetric vector potential of the Hulthén type [11]. They obtained the bound states energy spectrum and spinor wavefunctions in terms of the hypergeometric function.

In this article, we present a systematic development of the solution of this relativistic problem in the context of the Dirac equation using the same approach followed by Guo *et al.* However, we complement their solution by giving a complete one where the solution space splits into two distinct subspaces, one of which was obtained by the authors but not the other. The splitting depends on the sign of a fundamental parameter in the problem, which is a standard feature of the relativistic problem. For example, it is well-known that in relativistic problems with non zero angular momentum two regular solutions are obtained depending on the sign of $\kappa$, where $\kappa = \pm 1, \pm 2, ...$ for $\ell = j \pm \frac{1}{2}$. However, the current problem is analytically solvable only for $\ell = 0$. Nonetheless, we do find that in this case as well the solution space splits depending, however, on the sign of another dimensionless physical parameter symbolized by $\zeta$. The solution of the problem which was obtained in [11] is only for $\zeta < 0$ (in Ref. 11, the parameter $\lambda$ corresponds to $-\zeta$). Furthermore, the partial analysis of the energy spectrum which was given there for $\zeta = -1$ will be extended here to all values of $\zeta$. Interesting and unique properties of the spectrum will be pointed out and graphical representations will be given for several values of the physical parameters. The square integrable two-component spinor wavefunctions are written in terms of the Jacobi polynomials for all $\zeta$. Taking the nonrelativistic limit reproduces the nonrelativistic spectrum and gives an s-wave Schrödinger equation with a "generalized" three-parameter Hulthén potential, which is the sum of the original Hulthén potential and its square.

The paper is organized as follows. In Sec. II, we set up the three dimensional Dirac equation for a spinor coupled in a non-minimal way to the four-potential $(\mathcal{A}_0, \vec{\mathcal{A}})$. Spherical symmetry is imposed reducing the problem to a solution of the 2×2 radial Dirac equation involving two real radial functions, one of which is the independent potential function of the problem. A global unitary transformation is applied to the radial Dirac equation to separate the variables such that the resulting second order differential equation for the spinor components becomes Schrödinger-like. This requirement results in a linear constraint which relates the auxiliary radial function to the independent potential function. In Sec. III this scheme is implemented in the case where the potential function, which is the time component of the four-vector $(\mathcal{A}_0, \vec{\mathcal{A}})$, is taken as the Hulthén potential. The resulting s-wave Schrödinger-like equation is solved by comparison with that for an $L^2[-1,+1]$ function which is written in terms of the Jacobi polynomial. The solution of the problem (the upper and lower components of the spinor wavefunction and the relativistic bound states energy spectrum) is obtained for all values of the parameter $\zeta$. The paper concludes with a brief analysis of the energy spectrum.

## II. SOLUTION APPROACH FOR THE DIRAC EQUATION

Dirac equation is a relativistically covariant first order differential equation in four dimensional space-time for a spinor wave function $\psi$. For a free structureless particle and in the atomic units $\hbar = m = 1$, it reads $\left(i\gamma^\mu \partial_\mu - \lambdabar^{-1}\right)\psi = 0$, where $\lambdabar$ is the Compton wavelength $\hbar/mc = c^{-1}$. $m$ is the rest mass of the particle and $c$ is the speed of light. The summation convention over repeated indices is used. That is, $\gamma^\mu \partial_\mu \equiv \sum_{\mu=0}^{3} \gamma^\mu \partial_\mu = \gamma^0 \partial_0 + \vec{\gamma}\cdot\vec{\partial} = \lambdabar\gamma^0 \frac{\partial}{\partial t} + \vec{\gamma}\cdot\vec{\nabla}$. $\{\gamma^\mu\}_{\mu=0}^{3}$ are four constant square matrices satisfying the anti-commutation relation $\{\gamma^\mu, \gamma^\nu\} = \gamma^\mu \gamma^\nu + \gamma^\nu \gamma^\mu = 2\mathcal{G}^{\mu\nu}$, where $\mathcal{G}$ is the metric of Minkowski space-time which is equal to $\text{diag}(+,-,-,-)$. These are unimodular even dimensional matrices with a minimum dimension of four corresponding to spin ½ representation of the Lorentz space-time symmetry group. A four-dimensional matrix representation that satisfies the anticommutation relation is chosen as follows:

$$\gamma^0 = \begin{pmatrix} I & 0 \\ 0 & -I \end{pmatrix}, \quad \vec{\gamma} = \begin{pmatrix} 0 & \vec{\sigma} \\ -\vec{\sigma} & 0 \end{pmatrix} \tag{2.1}$$

where $I$ is the 2×2 unit matrix and $\vec{\sigma}$ are the three 2×2 hermitian Pauli matrices. Now, we let the Dirac particle be coupled to a four-component potential $\mathcal{A}_\mu = (\mathcal{A}_0, \vec{\mathcal{A}})$. Gauge invariant coupling is accomplished by the "minimal" substitution $\partial_\mu \to \partial_\mu + i\lambdabar\mathcal{A}_\mu$. This transforms the free Dirac equation to

$$\left[i\gamma^\mu(\partial_\mu + i\lambdabar\mathcal{A}_\mu) - \lambdabar^{-1}\right]\psi = 0 \tag{2.2}$$

where $\psi$ is a four-component spinor. When written in details, this equation reads

$$i\lambdabar\gamma^0 \frac{\partial}{\partial t}\psi = \left(-i\vec{\gamma}\cdot\vec{\nabla} + \lambdabar\vec{\gamma}\cdot\vec{\mathcal{A}} + \lambdabar\gamma^0 \mathcal{A}_0 + \lambdabar^{-1}\right)\psi \tag{2.3}$$

Multiplying both sides by $\lambdabar^{-1}\gamma^0$ gives

$$i\frac{\partial}{\partial t}\psi = \left(-i\lambdabar^{-1}\vec{\alpha}\cdot\vec{\nabla} + \vec{\alpha}\cdot\vec{\mathcal{A}} + \mathcal{A}_0 + \lambdabar^{-2}\beta\right)\psi \tag{2.4}$$

where $\vec{\alpha}$ and $\beta$ are the hermitian matrices

$$\vec{\alpha} = \gamma^0\vec{\gamma} = \begin{pmatrix} 0 & \vec{\sigma} \\ \vec{\sigma} & 0 \end{pmatrix} \text{ and } \beta = \gamma^0 = \begin{pmatrix} I & 0 \\ 0 & -I \end{pmatrix} \tag{2.5}$$

For time independent potential, Eq. (2.4) gives the following matrix representation of the Dirac Hamiltonian, in units of $mc^2 = 1/\lambdabar^2$,

$$H = \begin{pmatrix} \lambdabar^2 \mathcal{A}_0 + 1 & -i\lambdabar\vec{\sigma}\cdot\vec{\nabla} + \lambdabar^2 \vec{\sigma}\cdot\vec{\mathcal{A}} \\ -i\lambdabar\vec{\sigma}\cdot\vec{\nabla} + \lambdabar^2 \vec{\sigma}\cdot\vec{\mathcal{A}} & \lambdabar^2 \mathcal{A}_0 - 1 \end{pmatrix} \tag{2.6}$$

The wave equation reads $(H - \varepsilon)\psi = 0$, where $\varepsilon$ is the relativistic energy which is real and measured in units of $mc^2$. It should be noted that in our chosen units ($\hbar = m = 1$) the role of the fine structure constant is played by the Compton wavelength $\lambdabar$. The units ($\hbar = c = 1$) where the fine structure constant is used as the relativistic parameter are suitable for the electromagnetic interaction. The units that we are adopting here, where the relativistic parameter is $\lambdabar$, are suitable for dealing with a larger class of problems.

Equation (2.2) is invariant under the usual gauge transformation $\mathcal{A}_\mu \to \mathcal{A}_\mu + \partial_\mu \Lambda$, $\psi \to e^{-i\lambdabar\Lambda}\psi$, where $\Lambda(t,\vec{r})$ is a real space-time scalar function. Consequently, the off diagonal term $\lambdabar^2 \vec{\sigma}\cdot\vec{\mathcal{A}}$ in the Hamiltonian (2.6) could be eliminated ("gauged away") by a suitable choice of the gauge field $\Lambda(\vec{r})$. However, our choice of coupling will be non-minimal, which is obtained by the replacement $\lambdabar^2 \vec{\sigma}\cdot\vec{\mathcal{A}} \to \pm i\lambdabar^2 \vec{\sigma}\cdot\vec{\mathcal{A}}$, respectively. That is the Hamiltonian (2.6) is replaced by the following

$$H = \begin{pmatrix} \lambdabar^2 \mathcal{A}_0 + 1 & -i\lambdabar\vec{\sigma}\cdot\vec{\nabla} + i\lambdabar^2\vec{\sigma}\cdot\vec{\mathcal{A}} \\ -i\lambdabar\vec{\sigma}\cdot\vec{\nabla} - i\lambdabar^2\vec{\sigma}\cdot\vec{\mathcal{A}} & \lambdabar^2 \mathcal{A}_0 - 1 \end{pmatrix} \qquad (2.7)$$

It should be noted that this type of coupling does not support an interpretation of $(\mathcal{A}_0, \vec{\mathcal{A}})$ as the electromagnetic potential unless, of course, $\vec{\mathcal{A}} = 0$ (e.g., the Coulomb potential). Likewise, $H$ does not have local gauge symmetry. That is, the associated wave equation is not invariant under the electromagnetic gauge transformation mentioned above.

We impose spherical symmetry and write $(\mathcal{A}_0, \vec{\mathcal{A}})$ as $[V(r), \frac{1}{\lambdabar}\hat{r}W(r)]$, where $\hat{r}$ is the radial unit vector. $V(r)$ and $W(r)$ are real radial functions referred to as the even and odd components of the relativistic potential, respectively. In this case, the angular variables could be separated and we can write the spinor wavefunction as [12]

$$\psi = \begin{pmatrix} i[g(r)/r]\chi_{\ell m}^j \\ [f(r)/r]\vec{\sigma}\cdot\hat{r}\chi_{\ell m}^j \end{pmatrix} \qquad (2.8)$$

where $f$ and $g$ are real radial square-integrable functions. The angular component of the spinor wave-function is written as

$$\chi_{\ell m}^j(\hat{r}) = \frac{1}{\sqrt{2\ell+1}}\begin{pmatrix} \sqrt{\ell \pm m + 1/2}\; Y_\ell^{m-1/2} \\ \pm\sqrt{\ell \mp m + 1/2}\; Y_\ell^{m+1/2} \end{pmatrix}, \qquad \text{for } j = \ell \pm \tfrac{1}{2} \qquad (2.9)$$

$Y_\ell^{m\pm 1/2}(\hat{r})$ is the spherical harmonic function and $m$ stands for the integers in the range $-j, -j+1, \ldots, j$ and should not be confused with the mass. Spherical symmetry gives $i\vec{\sigma}\cdot(\vec{r}\times\vec{\nabla})\psi(r,\hat{r}) = -(1+\kappa)\psi(r,\hat{r})$, where $\kappa$ is the spin-orbit quantum number defined as $\kappa = \pm(j+\tfrac{1}{2}) = \pm 1, \pm 2, \ldots$ for $\ell = j \pm \tfrac{1}{2}$. Using this we obtain the following useful relations

$$\begin{aligned}(\vec{\sigma}\cdot\vec{\nabla})(\vec{\sigma}\cdot\hat{r})F(r)\chi_{\ell m}^j(\hat{r}) &= \left(\frac{dF}{dr} + \frac{1-\kappa}{r}F\right)\chi_{\ell m}^j \\ (\vec{\sigma}\cdot\vec{\nabla})F(r)\chi_{\ell m}^j(\hat{r}) &= \left(\frac{dF}{dr} + \frac{1+\kappa}{r}F\right)(\vec{\sigma}\cdot\hat{r})\chi_{\ell m}^j\end{aligned} \qquad (2.10)$$

Employing these in the wave equation $(H - \varepsilon)\psi = 0$ results in the following 2×2 matrix equation for the two radial spinor components

$$\begin{pmatrix} +1+\lambdabar^2 V(r)-\varepsilon & \lambdabar\left[\frac{\kappa}{r}+W(r)-\frac{d}{dr}\right] \\ \lambdabar\left[\frac{\kappa}{r}+W(r)+\frac{d}{dr}\right] & -1+\lambdabar^2 V(r)-\varepsilon \end{pmatrix}\begin{pmatrix} g(r) \\ f(r) \end{pmatrix}=0 \qquad (2.11)$$

This matrix equation results in two coupled first order differential equations for the two radial spinor components. Eliminating one component in favor of the other gives a second order differential equation. This will not be Schrödinger-like (i.e., it contains first order derivatives) unless $V = 0$. To obtain a Schrödinger-like equation in the general case we proceed as follows. A global unitary transformation $\mathcal{U}(\eta) = \exp(\frac{i}{2}\lambdabar\eta\sigma_2)$ is applied to the Dirac equation (2.11), where $\eta$ is a real constant parameter and $\sigma_2$ is the 2×2 Pauli matrix $\begin{pmatrix} 0 & -i \\ i & 0 \end{pmatrix}$. The Schrödinger-like requirement relates the two potential components by the linear constraint $V(r) = \xi[W(r) + \kappa/r]$, where $\xi$ is a real parameter. It also requires that $\sin(\lambdabar\eta) = \pm\lambdabar\xi$, where $-\frac{\pi}{2} < \lambdabar\eta < +\frac{\pi}{2}$. This results in a Hamiltonian that will be written in terms of only one arbitrary potential function; either the even potential component $V(r)$ or the odd one $W(r)$. Moreover, the solution of the problem is obtained for a given value of $\kappa$. It is to be noted that the angular parameter of the unitary transformation $\mathcal{U}(\eta)$ was intentionally split as $\lambdabar\eta$ and not collected into a single angle, say $\varphi$. This is suggested by investigating the constraint $\sin(\varphi) = \pm\lambdabar\xi$ in the nonrelativistic limit ($\lambdabar \to 0$) where we should have $\sin(\varphi) \approx \varphi = \pm\lambdabar\xi$. It also makes it obvious that in the nonrelativistic limit the transformation becomes the identity (i.e., not needed).

The unitary transformation together with the potential constraint map Eq. (2.11) into the following one, which we choose to write in terms of the even potential component

$$\begin{pmatrix} C-\varepsilon+(1\pm1)\lambdabar^2 V & \lambdabar\left(\mp\xi+\frac{C}{\xi}V-\frac{d}{dr}\right) \\ \lambdabar\left(\mp\xi+\frac{C}{\xi}V+\frac{d}{dr}\right) & -C-\varepsilon+(1\mp1)\lambdabar^2 V \end{pmatrix}\begin{pmatrix} \phi^+(r) \\ \phi^-(r) \end{pmatrix}=0 \qquad (2.12)$$

where $C = \cos(\lambdabar\eta) = \sqrt{1-(\lambdabar\xi)^2} > 0$ and

$$\begin{pmatrix} \phi^+ \\ \phi^- \end{pmatrix} = \mathcal{U}\psi = \begin{pmatrix} \cos\frac{\lambdabar\eta}{2} & \sin\frac{\lambdabar\eta}{2} \\ -\sin\frac{\lambdabar\eta}{2} & \cos\frac{\lambdabar\eta}{2} \end{pmatrix}\begin{pmatrix} g \\ f \end{pmatrix} \qquad (2.13)$$

Equation (2.12) gives the following equation for one spinor component in terms of the other

$$\phi^{\mp}(r) = \frac{\lambdabar}{C\pm\varepsilon}\left[-\xi\pm\frac{C}{\xi}V(r)+\frac{d}{dr}\right]\phi^{\pm}(r) \qquad (2.14)$$

While, the resulting Schrödinger-like wave equation becomes

$$\left[-\frac{d^2}{dr^2}+\left(\frac{C}{\xi}\right)^2 V^2 \mp \frac{C}{\xi}\frac{dV}{dr}+2\varepsilon V-\frac{\varepsilon^2-1}{\lambdabar^2}\right]\phi^{\pm}(r)=0 \qquad (2.15)$$

In the nonrelativistic limit ($\lambdabar \to 0$), $\varepsilon \approx 1+\lambdabar^2 E$ and $C \approx 1-\frac{1}{2}\lambdabar^2\xi^2$. Therefore, Eq. (2.14) shows that $\phi^+$ is the larger of the two relativistic spinor components (i.e., $\phi^+$ is the component that survives the nonrelativistic limit, whereas $\phi^- \sim \lambdabar\phi^+ \to 0$). Consequently, if we favor the upper spinor component then our choice of sign in the transformation parameter constraint is the top + sign. That is, we choose $\sin(\lambdabar\eta) = +\lambdabar\xi$.

In all relativistic problems that have been successfully tackled so far by this approach [13], Eq. (2.15) is solved by correspondence with well-known exactly solvable nonrelativistic problems. This correspondence results in a parameter map that relates the two problems. Now, if the nonrelativistic problem is exactly solvable, then using this parameter map and the known nonrelativistic energy spectrum one can easily obtain the relativistic spectrum. In fact, the relativistic extension of any known dynamical relationship in the nonrelativistic theory could easily be obtained by this correspondence map. The Green's function, which has prime significance in the calculation of relativistic processes, is such an example [14]. Moreover, the spinor wavefunction is also obtained from the nonrelativistic wavefunction using the same parameter map. In the following section this approach will be used in finding the solution of the Dirac equation with $V(r)$ taken as the Hulthén potential.

An alternative, but equivalent, approach to the one give above is to postulate the one-parameter two-component equation (2.12) as the relativistic wave equation and show that in the nonrelativistic limit ($\hbar \to 0$) the nonrelativistic problem is recovered. However, in this case, one cannot claim that the relativistic problem is a unique extension of the nonrelativistic one.

## III. THE DIRAC-HULTHÉN PROBLEM

For this problem we take $V(r) = -A/(e^{\omega r} - 1)$, where $\omega$ is the effective screening range of the potential. $A$ and $\omega$ are real and $\omega$ positive. This problem was partially solved for s-wave ($\ell = 0$) by Guo *et al.* using the same approach presented above [11]. Equation (2.15) for the upper spinor component gives

$$\left[ -\frac{d^2}{dr^2} + \frac{\rho(\rho+\omega)}{(e^{\omega r}-1)^2} + \frac{\rho\omega - 2\varepsilon A}{e^{\omega r}-1} - \frac{\varepsilon^2 - 1}{\hbar^2} \right] \phi^+(r) = 0 \qquad (3.1)$$

where $\rho = \tau\sqrt{1-(\hbar A/\tau)^2}$ and $\tau = -A/\xi$. If we choose the transformation parameter $\xi$ such that $\rho = -\omega$ then we obtain the solution for a pure s-wave Dirac-Hulthén problem. Nevertheless, we consider here the solution of a "generalized" three-parameter Dirac-Hulthén problem where $\rho$ is arbitrary and the potential consists of the sum of two terms: the original Hulthén potential and its square. Taking $x = 1 - 2e^{-\omega r}$ maps real space into a bounded one. That is, $r \in [0, \infty] \to x \in [-1, +1]$. A square integrable function (with respect to the measure $dr = \frac{1}{\omega}\frac{dx}{1-x}$) in this configuration space that is compatible with the domain of the wave operator and satisfies the boundary conditions could be written as

$$\phi_n^+(r) = a_n (1+x)^\alpha (1-x)^\beta P_n^{(\mu,\nu)}(x) \qquad (3.2)$$

where $P_n^{(\mu,\nu)}(x)$ is the Jacobi polynomial and $a_n$ is the normalization constant

$$a_n = \sqrt{\frac{\omega(2n+\mu+\nu+1)}{2^{\mu+\nu+1}} \frac{\Gamma(n+1)\Gamma(n+\mu+\nu+1)}{\Gamma(n+\mu+1)\Gamma(n+\nu+1)}} \qquad (3.3)$$

The real parameters $\alpha, \beta > 0$ and $\mu, \nu > -1$. Using the differential equation, differential formulas of the Jacobi polynomial [15], and $\frac{d}{dr} = \omega(1-x)\frac{d}{dx}$ we can write

$$\frac{d^2\phi_n^+}{dr^2} = \omega^2 \frac{1-x}{1+x}\left\{\left[-n\left(x+\frac{\nu-\mu}{2n+\mu+\nu}\right)\left(\frac{\mu-2\beta}{1-x}+\frac{2\alpha-\nu-1}{1+x}\right)-n(n+\mu+\nu+1)-\alpha(2\beta+1)\right.\right.$$
$$\left.\left.+\beta^2\frac{1+x}{1-x}+\alpha(\alpha-1)\frac{1-x}{1+x}\right]\phi_n^+ + 2\frac{(n+\mu)(n+\nu)}{2n+\mu+\nu}\left(\frac{\mu-2\beta}{1-x}+\frac{2\alpha-\nu-1}{1+x}\right)\frac{a_n}{a_{n-1}}\phi_{n-1}^+\right\} \quad (3.4)$$

To eliminate the off-diagonal representation (the $\phi_{n-1}^+$ term) we should take $\mu = 2\beta$ and $\nu = 2\alpha - 1$. Consequently, we obtain the following second order differential equation

$$\left\{-\frac{d^2}{dr^2}+\omega^2\alpha(\alpha-1)\left(\frac{1-x}{1+x}\right)^2-\omega^2\left[n(n+2\alpha+\mu)+\alpha(\mu+1)\right]\frac{1-x}{1+x}+\frac{\mu^2\omega^2}{4}\right\}\phi_n^+ = 0 \quad (3.5)$$

Comparing this with Eq. (3.1) results in the following parameter assignment

$$\alpha = \begin{cases} 1+\zeta & ,\tau \geq 0 \\ -\zeta & ,\tau < 0 \end{cases} \quad , \quad \mu_n = \begin{cases} \lambda_{n+1}^\zeta & ,\tau \geq 0 \\ \lambda_n^{-\zeta} & ,\tau < 0 \end{cases} \quad (3.6)$$

where $\zeta \equiv \rho/\omega$ and $\lambda_n^\zeta = -(n+\zeta)+(n+\zeta)^{-1}(\zeta^2+2\varepsilon A/\omega^2)$. Moreover, the comparison of the two equations gives, as well, the following quadratic (in $\varepsilon$) parameter relation

$$\varepsilon^2 = 1 - (\hbar\omega/2)^2 \mu_n^2 \quad , n = 0, 1, 2, ..., n_{max} \quad (3.7)$$

The energy spectrum is obtained as the set of two real solutions $\{\varepsilon_n^\pm\}_{n=0}^{n_{max}}$ of this quadratic equation, where $n_{max}$ is the maximum integer $n$ that yields real solutions for Eq. (3.7).

The upper radial spinor component is obtained by substituting the parameters of (3.6) into the wavefunction (3.2) giving

$$\phi_n^+(r) = \begin{cases} a_n^{\zeta+1}\sqrt{2^{\lambda_{n+1}^\zeta}}e^{-\omega\lambda_{n+1}^\zeta r/2}(1-e^{-\omega r})^{\zeta+1} P_n^{(\lambda_{n+1}^\zeta, 2\zeta+1)}(1-2e^{-\omega r}) & , \tau \geq 0 \\ a_n^{-\zeta}\sqrt{2^{\lambda_n^{-\zeta}}}e^{-\omega\lambda_n^{-\zeta} r/2}(1-e^{-\omega r})^{-\zeta} P_n^{(\lambda_n^{-\zeta}, -2\zeta-1)}(1-2e^{-\omega r}) & , \tau < 0 \end{cases} \quad (3.8)$$

The lower component of the spinor wavefunction is obtained by substituting this in the "kinetic balance" relation (2.14) which, in the $x$-coordinate, reads as follows

$$\phi^-(r) = \frac{\hbar\omega}{\varepsilon+\rho/\tau}\left[\frac{A}{\tau\omega}+\zeta\left(\frac{1-x}{1+x}\right)+(1-x)\frac{d}{dx}\right]\phi^+(r) \quad (3.9)$$

where $\varepsilon \neq -\rho/\tau$. Using the following recursion relations and differential formula satisfied by the Jacobi polynomials [15],

$$\left(\frac{1+x}{2}\right)P_n^{(\mu,\nu)} = \frac{n+\nu}{2n+\mu+\nu+1}P_n^{(\mu,\nu-1)} + \frac{n+1}{2n+\mu+\nu+1}P_{n+1}^{(\mu,\nu-1)} \quad (3.10a)$$

$$P_n^{(\mu,\nu)} = \frac{n+\mu+\nu+1}{2n+\mu+\nu+1}P_n^{(\mu,\nu+1)} + \frac{n+\mu}{2n+\mu+\nu+1}P_{n-1}^{(\mu,\nu+1)} \quad (3.10b)$$

$$(1-x^2)\frac{d}{dx}P_n^{(\mu,\nu)} = -n\left(x+\frac{\nu-\mu}{2n+\mu+\nu}\right)P_n^{(\mu,\nu)} + 2\frac{(n+\mu)(n+\nu)}{2n+\mu+\nu}P_{n-1}^{(\mu,\nu)} \quad (3.10c)$$

and writing the energy eigenvalues as $\varepsilon_n = \begin{cases} \varepsilon_{n+1}^\zeta, \tau \geq 0 \\ \varepsilon_n^{-\zeta}, \tau < 0 \end{cases}$ in correspondence with $\mu_n$, we obtain the following expressions for the lower spinor component when $\tau \geq 0$

$$\phi_n^-(r) = \frac{\hbar\omega a_n^{\zeta+1}}{\varepsilon_{n+1}^\zeta+\rho/\tau}\sqrt{2^{\lambda_{n+1}^\zeta}}e^{-\omega\lambda_{n+1}^\zeta r/2}(1-e^{-\omega r})^\zeta\left[\left(\frac{-\zeta+A/\tau\omega}{n+1+\zeta+\lambda_{n+1}^\zeta/2}+1\right)(n+2\zeta+1)\times\right.$$
$$\left. P_n^{(\lambda_{n+1}^\zeta, 2\zeta)}(1-2e^{-\omega r}) + \left(\frac{-\zeta+A/\tau\omega}{n+1+\zeta+\lambda_{n+1}^\zeta/2}-1\right)(n+1)P_{n+1}^{(\lambda_{n+1}^\zeta, 2\zeta)}(1-2e^{-\omega r})\right] \quad (3.11a)$$

For $\tau < 0$, the result is

$$\phi_n^-(r) = -\frac{2\lambdabar\omega a_n^{-\zeta}}{\varepsilon_n^{-\zeta}+\rho/\tau}\sqrt{2^{\lambda_n^{-\zeta}}}e^{-\omega\lambda_n^{-\zeta}r/2}(1-e^{-\omega r})^{-\zeta}\left\{\left(n+\lambda_n^{-\zeta}/2 - A/\tau\omega\right)P_n^{(\lambda_n^{-\zeta},-2\zeta-1)}(1-2e^{-\omega r}) + \right.$$
$$\left. -(1-e^{-\omega r})^{-1}\left[(2\zeta+1)P_n^{(\lambda_n^{-\zeta},-2\zeta-1)}(1-2e^{-\omega r}) + (n-2\zeta-1)P_{n+1}^{(\lambda_n^{-\zeta},-2\zeta-2)}(1-2e^{-\omega r})\right]\right\}$$
(3.11b)

Using the Jacobi polynomial identity $P_n^{(\mu,\nu)}(-1) = (-)^n \frac{\Gamma(n+\nu+1)}{\Gamma(n+1)\Gamma(\nu+1)}$ one can verify that the term inside square brackets in (3.11b) vanishes at $r = 0$. Moreover, one can also show that

$$\lim_{\delta\to 0^+}\left[\nu P_n^{(\mu,\nu)}(-1+\delta) - (n+\nu)P_n^{(\mu,\nu-1)}(-1+\delta)\right] = \frac{(-)^n\Gamma(n+\nu+1)}{2\Gamma(n)\Gamma(\nu+2)}(n+\mu)\,\delta + O(\delta^2) \quad (3.12)$$

Therefore, the term with the $(1-e^{-\omega r})^{-1}$ factor in (3.11b) is finite at $r = 0$ maintaining square integrability of $\phi_n^-(r)$.

We conclude with a brief analysis of the relativistic energy spectrum $\{\varepsilon_n^\pm\}_{n=0}^{n_{\max}}$. One can easily show that the two solutions of the quadratic equation (3.7) are:

$$\varepsilon_n^\pm = \left[1+\left(\tfrac{\lambdabar A/\omega}{n+\alpha}\right)^2\right]^{-1}\left\{\tfrac{\lambdabar^2 A}{2}\left[1-\left(\tfrac{\zeta}{n+\alpha}\right)^2\right]\pm\sqrt{1+\left(\tfrac{\lambdabar A/\omega}{n+\alpha}\right)^2 - \left(\tfrac{\lambdabar\omega}{2}\right)^2(n+\alpha)^2\left[1-\left(\tfrac{\zeta}{n+\alpha}\right)^2\right]^2}\right\} \quad (3.13)$$

The reality constraint on this energy spectrum requires that the expression under the square root be positive. This results in the condition that $n \le n_{\max}$, where $n_{\max}$ is the maximum integer that is less than or equal to

$$\left\{\zeta^2 + \frac{2}{(\lambdabar\omega)^2}\left[1+\sqrt{1+(\lambdabar\tau)^2}\right]\right\}^{1/2} - \alpha \quad (3.14)$$

where $\tau^2 = \zeta^2\omega^2 + \lambdabar^2 A^2$. The condition for the existence of bound states could then be translated into the condition that $n_{\max} \ge 0$. It is evident from (3.14) that $n_{\max} \to \infty$ in the nonrelativistic limit ($\lambdabar \to 0$) or in the no screening limit ($\omega \to 0$). The energy spectrum is made up of two halves: the upper half which is the collection of all energy eigenvalues in the set $\{\varepsilon_n^+\}_{n=0}^{n_{\max}}$, and the lower half which is made up of the set $\{\varepsilon_n^-\}_{n=0}^{n_{\max}}$. The lower bound of the upper half of the spectrum will either be $\varepsilon_{n_{\max}}^+$ or $\varepsilon_0^+$. On the other hand, $\varepsilon_{n_{\max}}^-$ or $\varepsilon_0^-$ will be the upper bound for the lower half. The choice of one or the other depends on the values of the physical parameters in the problem. For a spectrum which is highly symmetric around the zero energy line, the two energy bounds are $\varepsilon_{n_{\max}}^\pm$. Generally, for a symmetric spectrum, the two bounds match. That is, they go together either as $\varepsilon_{n_{\max}}^\pm$ or as $\varepsilon_0^\pm$. It is easy to verify that for $n = 0$ and when $\tau < 0$ Eq. (3.13) gives $\varepsilon_0^\pm = \pm\rho/\tau$. The four energy bounds $\{\varepsilon_0^\pm|_{\tau<0}, \varepsilon_{n_{\max}}^\pm|_{\tau\ge 0}\}$ correspond to non-degenerate states, while all others do not. This is so because Eq. (3.13) or, alternatively, Eq. (3.7) states that $\varepsilon_n^\pm|_{\tau\ge 0} = \varepsilon_{n+1}^\pm|_{\tau<0}$ for $n = 0,1,2,\ldots,n_{\max}-1$ and for all $\tau$. Figure 1 shows the relativistic energy spectrum for a given set of physical parameters and for several values of the dimensionless parameter $\zeta$. For large values of the potential strength $A$, the energy spectrum bends upward becoming less symmetric as illustrated in Fig. 2. By studying Eq. (3.7), one can easily show that in the nonrelativistic limit ($\lambdabar \to 0$) the following energy spectrum is obtained

$$E_n = -\frac{\omega^2}{8}\left(n+\alpha-\frac{\zeta^2+2A/\omega^2}{n+\alpha}\right)^2, \qquad n = 0,1,2,\ldots \quad (3.15)$$

which agrees with the nonrelativistic results obtained in Refs. [5]. Figure 3 is a numerical illustration showing that in the nonrelativistic limit $\varepsilon_n \to E_n$. Specifically, using $\varepsilon_n^\pm$ from formula (3.13) and $E_n$ from (3.15), we find that the smaller $\lambdabar$ the better the following approximation

$$\varepsilon_n^\pm \approx \pm\sqrt{1+2\lambdabar^2 E_n} \approx \pm\left(1+\lambdabar^2 E_n\right) \tag{3.16}$$

FIGURES CAPTION:

Fig. 1 shows the relativistic energy spectrum of the s-wave "generalized" Dirac-Hulthén problem for several values of the dimensionless parameter $\zeta$. The other physical parameters are taken (in arbitrary length units for $\omega$) as $A = 2.5\omega^2$ and $\lambdabar = 0.2\omega^{-1}$. The figure also shows that in this example $n_{max} = 6, 7, 9$ for $\zeta = 5, -4, 0$, respectively.

Fig. 2 shows the energy spectrum for the same system of Fig. 1 except for a substantial increase in the strength of the potential, where $A = 50\omega^2$. It is evident that the spectrum becomes less symmetric bending upward. Moreover, for this case $n_{max} = 8, 9, 11$, respectively.

Fig. 3 shows that the relativistic energy spectrum (red dots) approaches the nonrelativistic spectrum (black circles) as $\lambdabar$ becomes smaller and smaller (shown in units of $1/\omega$). The other parameters in this illustration were taken as $A = 2\omega^2$ and $\zeta = 3$.

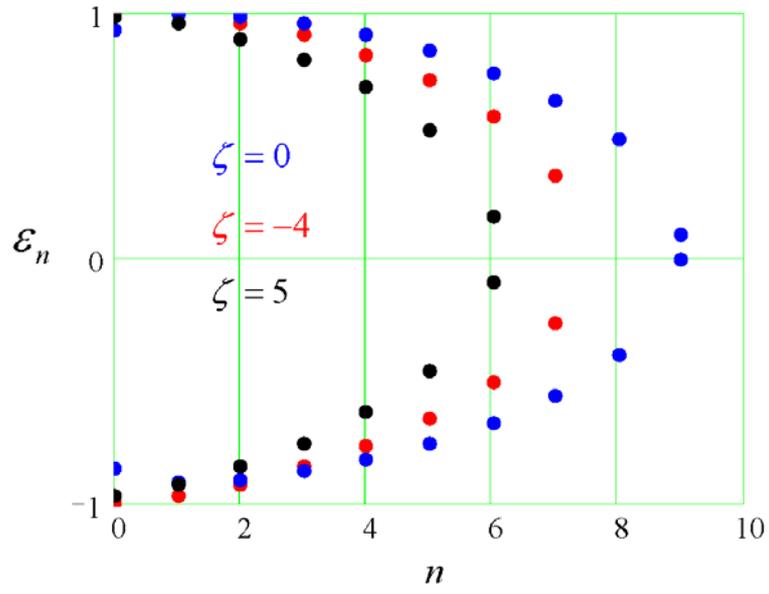

Fig. 1

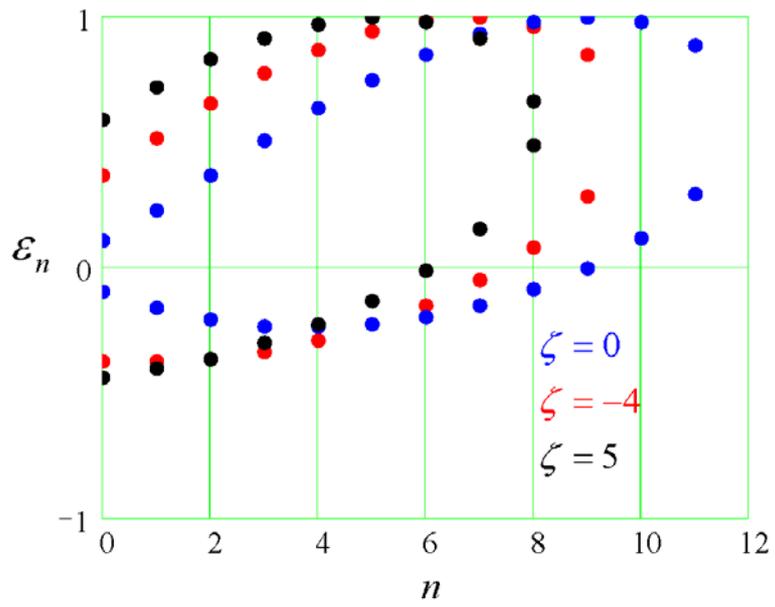

Fig. 2

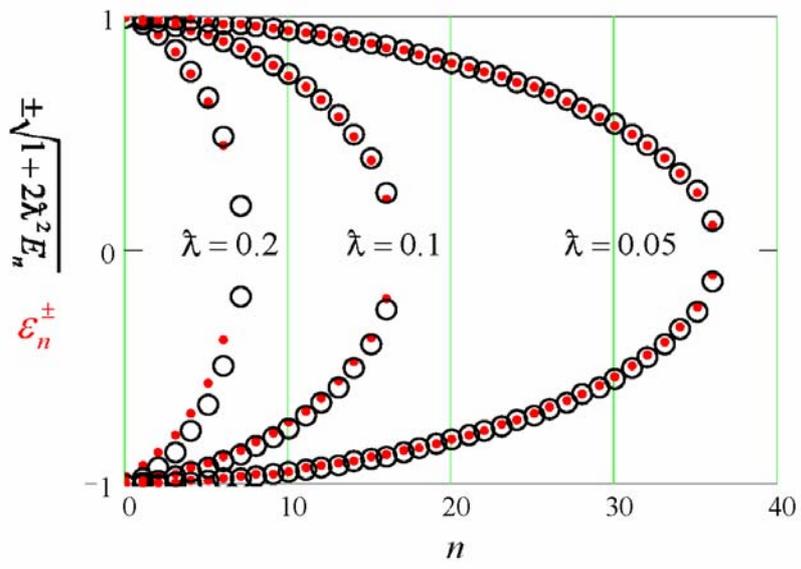

Fig. 3